\documentclass[showpacs,preprintnumbers,amsmath,floatfix]{revtex4}
\usepackage{graphicx}
\usepackage{dcolumn}
\usepackage{bm}

\begin{document}

%
\title{Bounds on Low Scale Gravity from RICE Data and Cosmogenic Neutrino Flux Models}

\author{Shahid Hussain}
\affiliation{Bartol Research Institute, University of Delaware, Newark, DE 19716}
\author{Douglas W. McKay}
\affiliation{University of Kansas Dept. of Physics and Astronomy, Lawrence KS
66045-2151, USA}

\date{\today}

\begin{abstract}

We explore limits on low scale gravity models set by results from the Radio Ice Cherenkov Experiment's (RICE) ongoing search for cosmic ray neutrinos in the cosmogenic, or GZK, energy range. The bound on  $M_{D}$, the fundamental scale of gravity, depends upon cosmogenic flux model, black hole 
formation and decay treatments, inclusion of graviton mediated elastic neutrino processes, and the number of large extra dimensions, d.  Assuming proton-based cosmogenic flux models that cover a broad range of flux possibilities,  we find bounds in the interval 0.9 TeV $<M_{D}<$ 10 $TeV$.  Heavy nucleus-based models generally lead to smaller fluxes and correspondingly weaker bounds.  Values d = 5, 6 and 7, for which laboratory and astrophysical bounds on LSG models are less restrictive, lead to essentially the same limits on $M_{D}$.

\end{abstract}

\maketitle

\section{Introduction}

The highest energy cosmic ray events range to values above $10^{5}PeV$ \cite
{agasahires}, with hundreds of $TeV$ equivalent center-of-mass energies.
Prospects of achieving comparable energies at an accelerator laboratory are
remote, at best. The opportunities for probing fundamental physics at these
ultrahigh energies will be confined to cosmic rays for the foreseeable
future. The weakness of neutrino interactions at currently available
energies allows neutrinos to carry information to us line of sight from deep
within highly energetic astrophysical sources. For the same reason,
enhancements of neutrino interaction rates by new physics at ultra-high
energies can be dramatic and distinctive. On the other hand, similar sized
enhancements in the electromagnetic and strong interaction rates can easily
be obscured. These two, unique features of neutrinos - direct propagation
from source to observer and sensitivity to new physics - drive the effort to
build neutrino telescopes to cover energies from fractions of an $MeV$ to
detect $p-p$ neutrinos from the sun to thousands of $PeV$ to detect
cosmogenic neutrinos, those originating from the decays of pions created by
collisions of the highest energy cosmic rays with the cosmic microwave
radiation photons \cite{gzk,berez}.

In this paper we present a detailed study of bounds on new physics,
specifically on the scale $M_{D}$ in models of low scale gravity \cite{ADD},
that follow from predictions by ultrahigh-energy (UHE) proton-based models
of cosmogenic neutrino fluxes \cite{ess,kkss,pj,frt} and the absence of UHE
neutrino events in the Radio Ice Cherenkov Experiment (RICE) data from 2000
through 2004 \cite{rice05}. We briefly compare these limits to limits
obtained by using predictions by recently proposed heavy nuclei-based models
for the cosmogenic neutrino flux \cite{nunucl}.

Before proceeding to a description of the calculation, the results and the
summary and conclusions, we sketch the essential features of the RICE
detection system and the features of low scale gravity that lead to
prediction of event rates in RICE that exceed those expected from neutrino
interactions in the standard model (SM).

\textbf{The RICE experiment} The advantages of weakly interacting neutrinos
raise the obvious question of detection. Neutrino detection is possible with
some combination of large fluxes and large detector volumes. Accelerator
neutrino facilities strive for high fluxes and detectors the size of large
fixed target experiments. But in the energy range above a $PeV$, the
predicted cosmogenic fluxes are extremely small and decrease rapidly with
energy. the high energy end of the expected cosmogenic flux range, $%
10^{4}-10^{5}PeV$, can be characterized by the observed cosmic ray figure of
1 event per $km^{2}$ per century. \textit{Many} cubic kilometers is the
effective detector volume that must be achieved.

The RICE detector concept relies on characteristics of UHE showers, radio
wavelength emission from these showers, and the transmission of radio
wavelengths in cold, pure ice \cite{askary}. The showers in dense media
(e.g. ice), are compact, travel faster than light in the medium and are
smaller in transverse size than sub-GHz frequency wavelengths. The showers
develop a net excess charge at shower maximum that is about $10^{6}$
electrons at a $PeV$, and is proportional to energy. This net charge emits 
\textit{coherent} Cherenkov radiation at frequencies up to a GHz. This
radiation has an attenuation length of more than a kilometer in Antarctic
ice. A single radio antenna can be sensitive to neutrino induced showers
within more than a $km^{3}$ of ice at the highest energies; even a modest,
pilot array has an effective volume, $V_{eff}$, of many cubic kilometers.
Discussion of the recent RICE analysis can be found in Ref. \cite{rice05},
while more details of the full experiment are given in Refs. \cite
{riceastro03} and \cite{riceicrc03}.

\textbf{Low scale gravity and UHE Cosmic Ray Neutrinos}

The phenomenology of low scale gravity (LSG) \cite{ADD} predicts enhanced
cross sections mediated by gravity, including production of black holes in
high energy particle reactions \cite{banks}, \cite{arg}. Under quite general
assumptions, the gravity enhanced cross sections, such as that for black
hole production, are large, and grow with a power of the center of mass
energy. After formation, black holes are thought to evaporate to a
statistical mixture of many particles \cite{hawking}, producing a
characteristic signature. There is great interest in black hole production
at current and future colliders as a ``low energy'' probe of extra
dimensions \cite{dimgid,cheung}. Yet even the LHC energies may not be large
enough to reach the black hole regime \cite{cheung}.

A number of ultra-high energy neutrino studies take advantage of the fact
that the highest energies accessible in particle physics occur not at
colliders, but in cosmic rays; this is then coupled to the prediction that
low scale gravity cross sections in extra dimension models, while much
smaller than hadronic, can be orders of magnitude larger than standard model
(SM) \textit{neutrino} cross sections at ultra-high energies\cite{lsgphen}.
With the assumption that the collision energy dumped into a black hole is
released as visible energy in air shower detectors or neutrino telescope
detectors, predictions of event rates and consequent bounds on $M_{D}$, the
``Planck mass'' of extra-dimension gravity, have been presented for various
model estimates of neutrino flux \cite{bhphen}.

A low scale gravity effect that does not depend on the black hole dynamics
is graviton exchange, producing an hadronic shower from the graviton -
parton interaction in the nucleon. This is a variety of deep-inelastic,
neutral-current process, and we include this in our study as well. This
class of events is complementary to black hole production and decay. Though
the cross sections are large, the elasticity is high, greater than $\sim
90\% $, so relatively little of the neutrino energy is deposited in the
showers. Nonetheless, in part of the parameter space this ``deep inelastic''
gravity effect competes with the black hole formation.

\section{Summary of direct black hole production and graviton meditated
neutrino deep inelastic scattering}

We present the essential components of our calculation of the number of
events expected in RICE data \cite{rice05} from SM, black hole, and
graviton-mediated interactions . For impact parameters less than the
Schwartzschild radius, $r_{S}$, the black disk, black hole cross section
formula has often been adopted, which reads 
\begin{equation}
\hat{\sigma}_{\mathrm{BH}}\approx \pi r_{S}^{2},  \label{sigma_BH}
\end{equation}
without including gray body \cite{graybody}, or impact parameter\cite
{impactpar} and form factor \cite{formfactor}, effects. Later we include a
model for the impact parameter dependence for comparison. In Eq. (1), $r_{S}$
is the 4+$d$ dimensional Schwartzschild radius of a black hole of mass $%
M_{BH}$ \cite{meyperr}: $r_{S}={\frac{1}{M_{D}}}\left[ \frac{M_{\mathrm{BH}}%
}{M_{D}}\right] ^{\frac{1}{1+d}} \kappa_{d}$, where $\kappa_{d}$ = 2.1,
2.44, or 2.76 for cases d = 5, 6, or 7, respectively. Here $M_{D}$ is the 4+$%
d$ dimensional Planck mass scale of extra-dimensional physics, which is the
scale we aim to explore with UHE RICE data in this study. Equation (1) is a
parton level cross section, and the effective black hole mass is $M_{BH}=%
\sqrt{\hat{s}}$, with $\hat{s}=xs$; s is the square of the center-of-mass
energy and $x$ is the momentum fraction of the struck parton. The
corresponding cross section is given by

\begin{equation}
\sigma _{\nu N\rightarrow BH}(E_{\nu})=\int_{x_{min}}^{1}dx\hat{\sigma}%
_{BH}(xs)\sum_{i}f_{i}(x,Q),
\end{equation}
where the $f_{i}(x,Q)$ are the parton distribution functions, $x_{\min
}=M_{BH0}^{2}/s$ or $1/r_{s}^{2}s$, whichever is larger; the sum over index $%
i$ accounts for different parton flavors. Here $s=2M_{N}E_{\nu}$ and $Q$ is
chosen to be $\sqrt{xs}$. Choosing $Q=\frac{1}{r_{s}}$ \cite{dimgid} makes
insignificant difference. $E_{\nu }$ is the primary neutrino energy. The
choices of $x_{min}$ and $Q$ are not unique, as discussed in the papers
listed in Refs. \cite{dimgid,cheung,bhphen}. Here $M_{BH0}$ is a free
parameter corresponding to the minimum energy needed to form a black hole.

So far we have described the simplest versions of BH\ formation to estimate
cross sections and UHE shower rates. Many possible modifications show up
when the horizon forming collision is studied in detail. A proposal by
Yoshino and Nambu \cite{impactpar} to sharpen the horizon formation
criterion is straightforward to implement \cite{afgs}, and we include it
here to illustrate the effect on the event rates for BH production. The
impact parameter dependence of the apparent horizon formation is
conveniently presented as a plot of mass contained within an apparent
horizon ($M_{A.H.}$) vs. impact parameter. We fit the graph of $M_{A.H.}$ vs 
$b$ with a function $M_{A.H.}(b/b_{\max }=z)$, the mass contained within an
horizon as a function of impact parameter, scaled by the impact parameter
beyond which no horizon forms, $b_{\max }$, as in Ref. \cite{impactpar}. We
set the lower limit of x, x$_{\min }=(M_{BH0})^{2}/M_{A.H.}^{2}(z).$ The
form of the cross section is still taken to be $\sigma (\hat{s})=\pi
r_{s}^{2}(\hat{s})$, but the threshold for BH production is raised, which
accounts for c.m. energy that does not contribute to BH formation and
subsequent decay to observable particles. In addition to the x-integral, the
cross section now includes an average over impact parameter, taken to be
weighted geometrically by the area element $d(\pi b^{2})/\pi b_{\max }^{2}$ 
\cite{afgs}. The cross section for this modeling of the impact-parameter
effects reads

\begin{equation}
\sigma _{\nu N\rightarrow BH}(E_{\nu })=\int_{0}^{1}2zdz\int_{x_{min}}^{1}dx%
\hat{\sigma}_{BH}(xs)\sum_{i}f_{i}(x,Q).
\end{equation}
Again, $s=2M_{N}E_{\nu }$ and $E_{\nu }$ is the primary neutrino energy. All
of the cross sections relevant in this paper are shown in Figure 1 for
representative values of $M_{D}$ and $M_{BH0}$, with $d=6$. The black hole
production cross sections are sensitive to the values of $M_{D}$ and $%
M_{BH0} $ and to the inclusion of impact parameter effects. The bounds we
obtain for $d=5$ or 7 differ by only five to ten percent from those obtained
for $d=6$, so we give details only for $d=6$ throughout. Laboratory and
astrophysical data already place strong lower bounds on the cases $d<5$.

We treat the graviton exchange in the higher dimension, low scale gravity
picture in the eikonal approximation \cite{many1}. For more than 3 spatial
dimensions, an impact parameter scale $b_{c}$ enters the problem, and the
dominant contribution to the eikonal amplitude when $\sqrt{\hat{s}}\gg M_{D}$
comes from momentum transfer $Q=\sqrt{-(p-p^{\prime })^{2}}$ in the range $%
1/r_{S}>Q>1/b_{c}$, where the stationary phase evaluation of the amplitude
is a good approximation. The four momenta $p$ and $p^{\prime }$ refer to the
incident and scattered neutrinos respectively. The eikonal amplitude can be
written in this approximation

\begin{equation}
|\mathcal{M}_{d}|=B_{d}(b_{c}M_{D})^{d+2}\left[ b_{c}q\right] ^{-\frac{d+2}{%
d+1}}\ ,
\end{equation}
where $B_{d}$ = 0.23, 0.039 or 0.0061 when $d=5$,6, or 7. The saddle point
impact parameter reads $b_{c}=\frac{1}{M_{D}}\left( {\frac{s}{M_{D}^{2}}}%
\right) ^{\frac{1}{d}}\beta _{d}$. The factor $\beta _{d}$ = 1.97, 2.32, or
2.66 for $d=5$, 6, or 7. The elastic parton level cross section then reads,

\[
\sigma _{EK}(x,q)=\frac{1}{16\pi xs}\sum_{i}f_{i}(x,q)|\mathcal{M}_{d}|^{2}, 
\]
which enters into the rate calculations below. The cross sections are
displayed in Fig.1 for the case $M_{D}$ = 1 $TeV$.

\begin{figure}[tbp]
\includegraphics[width=3.6 in, angle= 0 ]{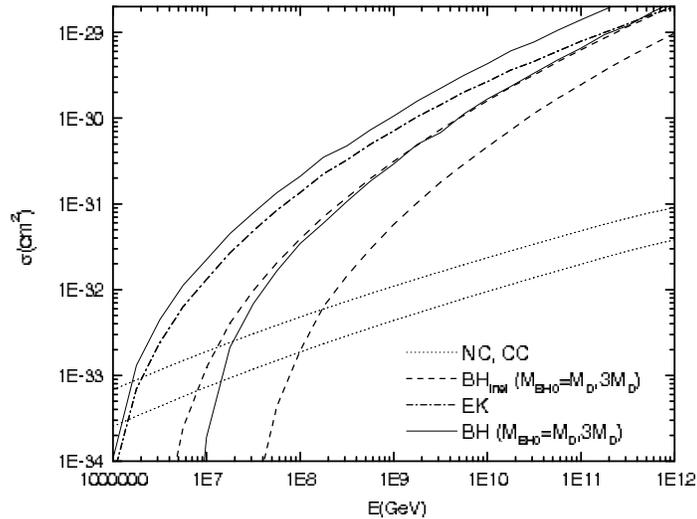}
\caption{Standard model and low scale gravity ($M_D$=1TeV, $d$=6) cross
sections. Pair of dotted curves gives standard model NC (lower curve) and CC
(upper curve) interactions. Pair of solid (dashed) curves gives black hole
formation cross section, without (with) inelasticity, for $M_{BH0}$=$M_D$
(upper solid (dashed) curve) and $M_{BH0}$=3$M_D$ (lower solid (dashed)
curve). The dash-dot curve is for graviton exchange case.}
\label{fig:xseclsg1}
\end{figure}

At this point we can write the shower production rates for the eikonal and
direct black hole production cases, specifying the restrictions on the range
of integrations from impact parameter and threshold considerations. For the
black hole production we have the folding of the neutrino flux energy
distributions with the effective volume of the detector and the scattering
cross sections on the nucleon target integrated over x. The eikonal set-up
has an additional integration over the inelasticity parameter $y$. These
expressions are presented below. These results have appeared in various
forms elsewhere \cite{prop}.

Shower rates from $BH$ production and decay, with or without including the
impact parameter effects (black hole inelasticity), are determined by
folding the appropriate cross section with the effective volume and
differential flux :

\begin{equation}
R_{shower}^{BH}=2\pi \rho N_{A}\sum_{i}\int\limits_{E_{th}}^{E_{\nu }\max
}dE_{\nu }\frac{dF_{i}(E_{\nu })}{dE_{\nu }}V_{eff}(E_{\nu })\sigma _{\nu
N\rightarrow BH}(E_{\nu }),
\end{equation}
where $x_{\min }=M_{BH0}^{2}/s$ or $1/r_{s}^{2}s$, whichever is larger, for
the simple black disk case (Eq. 2); $x_{min}=(M_{BH0})^{2}/M_{A.H.}^{2}(z)$
when the impact-parameter-dependent inelasticity estimate is included (Eq.
3). The mass enclosed in the apparent horizon, $M_{A.H.}$, and the scaled
impact parameter, $z$, are discussed above. $E_{th}$ is the experimental
threshold for detection of showers, taken to be $100PeV$ for RICE. Above the
cutoff $E_{\nu }\max $, the integral gives negligible contribution to R. The
rate shown is for down-going neutrinos, assuming an isotropic, diffuse flux
of neutrinos, appropriate to those of cosmogenic origin. Fluxes are given
per steradian, and the $2\pi $ factor accounts for the integration of
isotropic flux times angular average of the effective volume over the dome
of the sky.

For the eikonal scattering case, the rate expression reads

\begin{equation}
R_{shower}^{EK}=2\pi \rho N_{A}\sum_{i}\int\limits_{E_{th}}^{E_{\nu }\max
}dE_{\nu }\frac{dF_{i}(E_{\nu })}{dE_{\nu }}\int\limits_{E_{th}/E_{\nu
}}^{1}dyV_{eff}(yE_{\nu })\int\limits_{M_{D}^{2}/ys}^{1/r_{s}^{2}ys}dx\sigma
_{EK}(x,q),
\end{equation}
with $q=\sqrt{xys}$. The graviton exchange is affected only by the value of $%
M_{D}$, though there is an implicit dependence on the choice of minimum
impact parameter where the eikonal, semiclassical approximation is expected
to be valid. We have chosen this value as $1/r_{s}$. The fluxes we used to
explore the range of event rates are shown in Fig. 2. Recent work that
estimates cosmogenic neutrino fluxes based on UHE nuclei is reported in Ref. 
\cite{nunucl}.

\begin{figure}[tbp]
\includegraphics[width=3.6 in, angle = 0] {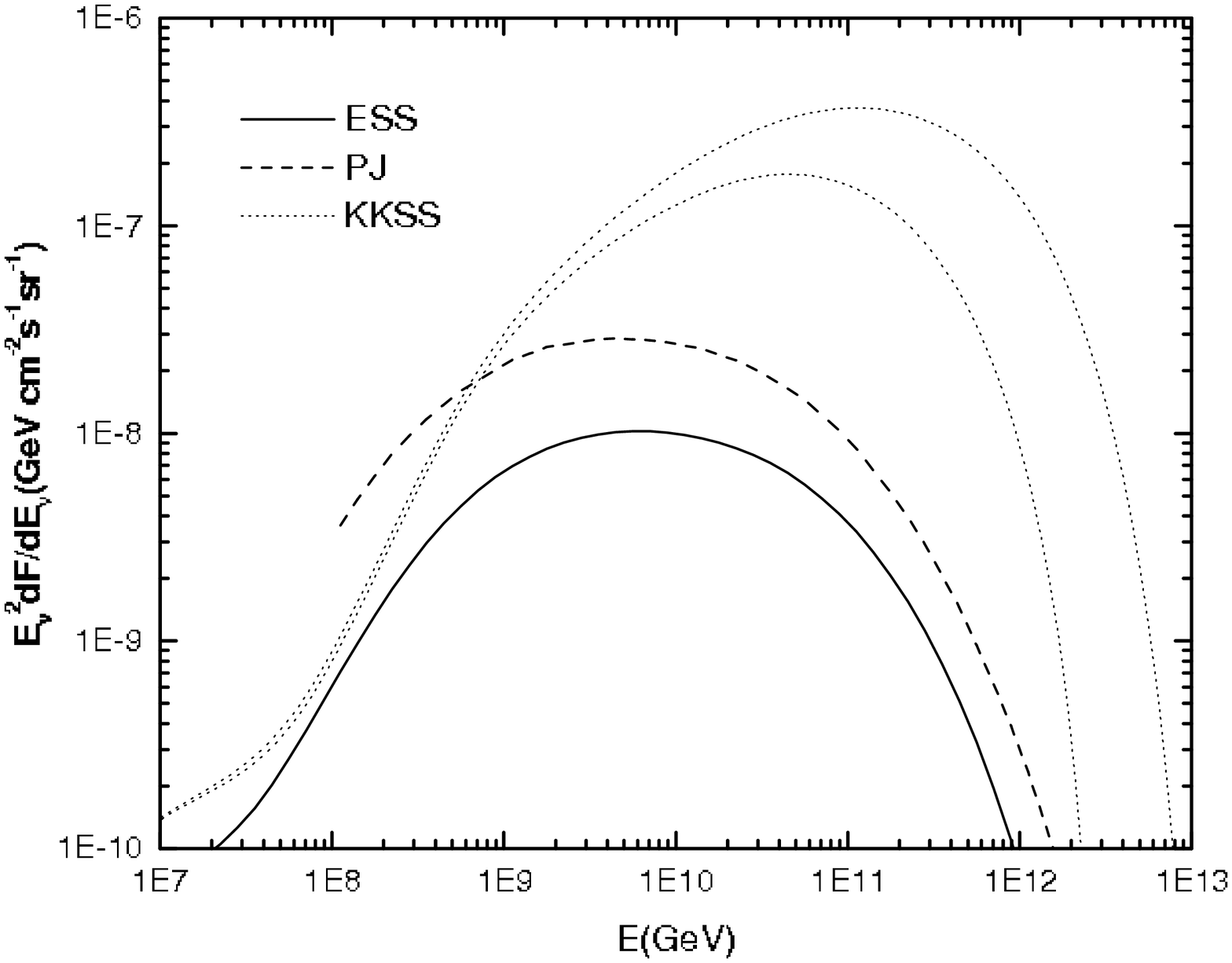} \vspace{0.0 cm} ere is
how to import EPS art
\caption{Cosmogenic $\nu _{e}$ fluxes by ESS \protect\cite{ess}, PJ 
\protect\cite{pj}, and KKSS \protect\cite{kkss}. We show the KKSS maximal
and a relatively smaller flux for the KKSS case.}
\label{fig:gzknu}
\end{figure}

\section{Numerical work and results}

To produce representative event rate predictions, we fix number of extra
dimensions $d$ at 6, consider both $M_{D}=1TeV$ and $2TeV$ and the minimum
BH mass values $M_{BH0}=M_{D}$ and $3M_{D}$. As we noted in our outline of
cross sections, changing $d$ to 5 or 7 affects our bounds by only 5 - 10
percent. We use CTEQ6 \cite{cteq} for the nucleon parton distributions. The
cross sections for $M_{D}=1TeV$ are plotted from $10^{6}GeV$ to $10^{12}GeV$
in Fig. 1.

When $M_{D}$ = 2 $TeV$, each cross section is reduced by roughly a factor
10. To assess flux model dependence on results, we chose large, medium, and
small fluxes for comparison, as described in the caption of Fig. 2. We are
primarily interested in cosmogenic flux models, generated by the GZK effect 
\cite{ess,kkss,pj,frt}. These produce neutrinos at the very highest
energies, and are arguably less model-dependent than those generated from
models of specific types of astrophysical sources. In Fig. 2 we show the
comparison between the three proton-based flux models used, which show
factors of three or more difference over most of the energy range of
interest, and more than an order of magnitude in some regions \cite
{ess,kkss,pj,frt}. Though this and previous studies, for example \cite
{bhphen, prop}, focus on the proton-based cosmogenic flux model, we also
comment below on the consequence if heavy nuclei \cite{nunucl} are the UHE
cosmic rays rather than protons. These latter models themselves vary widely
in their neutrino flux predictions, depending upon assumed energy cutoff of
the source and on nuclear species. Moreover, a mixture of several species,
including protons, is consistent with cosmic ray data.

The qualitative features of our calculations in all channels can be
anticipated from the power law growth of cross sections, times stronger
power law fall of flux, times (slower) power law growth of effective volume
to detect shower energy. The rapid decrease in flux with energy wins, but
the growing cross sections and effective volume delay the loss of events as
one scans higher and higher energies. In Table I we present the number of
events expected in RICE over the 5 year period 2000-2004 using the three
different cosmogenic flux models. We have integrated using an updated
effective volume of the RICE detector \cite{rice05} for our visible energy
signal. In all cases $E_{th}$ (our detector threshold) is $10^{7}$ $GeV$.
All rates are integrated over primary neutrino energy from threshold to the
maximum energy where the flux cuts off. As shown in Table I, since at 95\%
C.L., with SM cross sections, the maximum number of events compatible with
no background and zero observed events is 3.0, RICE nearly rules out the
maximal KKSS cosmogenic flux model. The $M_{D}=1$ and $d=6$ choices made for
Table I are irrelevant for the SM result, of course.

For the SM, EK and the direct BH production cases, the hadronic component of
showers is assigned pulse strength equal to an electromagnetic shower of the
same energy \cite{shshow}. An important difference between electromagnetic
and hadronic showers is that the hadronic showers do not suffer the loss of
effective volume that is estimated to affect the electromagnetic showers due
to the LPM \cite{lpm} effect when computing event rates.

\begin{widetext}
\begin{center}
\begin{table}
\caption{ $95\%$ C.L. upper limit shower rates, for RICE 2000-2004 operation, 
in low scale gravity ($M_{D}=1$ TeV, $d=6$); rates are given for different cosmogenic
neutrino flux models; $BH$ and $BH_{inel}$ give black hole formation rates
(and hence shower rates) without and with inelasticity included,
respectively; EK gives the shower rates due to graviton exchange. KKSS corresponds to 
the maximal flux case shown in figure 2.}
\label{tab:tbl3}
\begin{tabular}{|c|c|c|c|c|}
\hline
Flux & $SM$ & $EK$ & $BH1$ ($M_{BH_{0}}$=$M_{D}$) & $BH2$ ($M_{BH0}$=3$M_{D}$)
\\ \hline
$
\begin{array}{c}
\\ 
ESS \\ 
\\ 
KKSS \\ 
\\ 
PJ
\end{array}
$ & $
\begin{array}{c}
\\ 
0.1 \\ 
\\ 
3.0 \\ 
\\ 
0.27  
\end{array}
$&$
\begin{array}{c}
\\ 
0.66 \\ 
\\ 
43 \\ 
\\ 
1.7
\end{array}
$ & $
\begin{array}{c}
BH \\ 
38.4 \\ 
\\ 
1509 \\ 
\\ 
106
\end{array}
\begin{array}{c}
BH_{inel} \\ 
4.0 \\ 
\\ 
238 \\ 
\\ 
11
\end{array}
$ & $
\begin{array}{c}
BH \\ 
15 \\ 
\\ 
697 \\ 
\\ 
41
\end{array}
\begin{array}{c}
BH_{inel} \\ 
1.4 \\ 
\\ 
98 \\ 
\\ 
3.7
\end{array}
$ \\ \hline
\end{tabular}
\end{table}
\end{center}
\end{widetext}

In the EK, BH1, and BH2 columns of Tables I and II, we show expected numbers
of events in the RICE 2000-2004 data, with the acceptance (effective volume)
and efficiencies in candidate event analysis included, coming individually
from the eikonalized graviton exchange (EK), and black hole production
processes. Representative scales $M_{D}=1,2$ $TeV$ are used, and for each
the black hole formation thresholds $M_{BH0}=M_{D},$ $3M_{D}$ are assumed
and the black hole rate calculated. It is usually argued \cite{banks,dimgid}
that in the range $M_{D}<$ $M_{BH0}<2M_{D}$, the quantum gravity effects may
dominate \cite{edgrw, mfairb}; our point of view is that the new physics
leading to shower formation will occur, and the EK and BH shower rates act
as reasonable estimates of the enhancement in the transition region.
Certainly one expects that the different energy regions tie together
smoothly parametrically, and they do \cite{edgrw}. In addition we show two
values of black hole events for each black hole $M_{D},$ $M_{BH0}$ parameter
pair, corresponding to the simple ``black disk'' calculation and the
calculation corrected for loss of efficiency for black hole formation when
the impact parameter offset between neutrino and parton increases. The
second number in the columns labeled ``$BH_{inel}$'' is that obtained when
the inelasticity effects are modeled as described above. The black hole ($BH$%
) rates drop by an order of magnitude. The EK rate, unaffected by the $BH$
inelasticity effects, is now about 20\% of the $BH_{inel}$ for $M_{BH}=M_{D}$
case and 50\% of $BH_{inel}$ for the $M_{BH}=3M_{D}$ case. Again this is
true for all the flux models we apply, which cover a wide range of
possibilities.

Pumping up the cross sections above SM extrapolations pushes harder on the
flux bounds. Given that UHE neutrinos are not yet observed \cite{rice05},
choosing a flux model then allows one to limit cross sections, which predict
at some point more events than current observational limits allow. For RICE,
this number is 3.0 at 95\% C.L., as noted earlier. Clearly most of the
entries in Tables I and II are already disallowed at better than 95\% C.L.
To quantify this situation, we allow the mass scale of gravity to decrease
(increase) and ask what value allows the event rate for a given flux to rise
above (fall below) the 95\% limit set by RICE. This value is what we quote
in Table III and shown as a function of $M_{BH0}/M_{BH}$ in Fig. 3.

To assess the effect of BH inelasticity on $M_{D}$ bounds, we also use the
impact parameter dependent cross sections as modeled in \cite{impactpar,afgs}%
, to place bounds on $M_{D}$ for each flux and $M_{BH0}$ choice. In Table
III the $BH$ and $ALL$ columns again show two values each. The first refers
to the case where all parton CM collision energy is available to the BH,
while the second is the value after the modeling of inelasticity is
included. The reduction in sensitivity to $M_{D}$ is substantial, with
reduction ranging from factors of 3/4 to 1/2.

\begin{widetext}
\begin{center}
\begin{table}
\caption{Same as above with $M_{D}=2TeV$.}
\label{tab:tbl4}
\begin{tabular}{|c|c|c|c|}
\hline
Flux & $EK$ & $BH1$ ($M_{BH_{0}}$=$M_{D}$) & $BH2$($M_{BH_{0}}$=3$M_{D}$) \\ 
\hline
$
\begin{array}{c}
\\ 
ESS \\ 
\\ 
KKSS \\ 
\\ 
PJ
\end{array}
$ & $
\begin{array}{c}
\\ 
0.10 \\ 
\\ 
7.2 \\ 
\\ 
0.26
\end{array}
$ & $
\begin{array}{c}
BH \\ 
4.5 \\ 
\\ 
195 \\ 
\\ 
12
\end{array}
\begin{array}{c}
BH_{inel} \\ 
0.30 \\ 
\\ 
29 \\ 
\\ 
1.2
\end{array}
$ & $
\begin{array}{c}
BH \\ 
1.5 \\ 
\\ 
78 \\ 
\\ 
4.0
\end{array}
\begin{array}{c}
BH_{inel} \\ 
0.12 \\ 
\\ 
10 \\ 
\\ 
0.32
\end{array}
$ \\ \hline
\end{tabular}
\end{table}
\end{center}
\end{widetext}

\begin{widetext}
\begin{center}
\begin{table}
\caption{Experimental lower bounds on LSG scale $M_{D}$, based on 2000-2004 RICE data
(0 events, 0 background). Here `$ALL$' is the combined bound due to $EK$ and 
$BH$; these bounds are due to all flavors. The pairs of numbers under
columns $BHD$ and `$ALL$' are the bounds without and with black hole
inelasticity, respectively. The bounds here include SM interactions. The
numbers are in TeV. KKSS corresponds to the 
maximal flux case shown in figure 2. We fix number of extra dimensions d to 6. With d = 5 or 7, 
we find values 2.0 TeV or 2.3 TeV instead of 2.15 TeV for the first entry in the ESS, BH slot in the table.  
These are typical fractional changes.}
\label{tab:tbl5}
\begin{tabular}{|c|c|c|}
\hline
Flux & $M_{BH_{0}}=M_{D}\text{ }$ & $M_{BH_{0}}=3M_{D}$ \\ \hline
$
\begin{array}{c}
\\ 
ESS \\ 
\\ 
KKSS \\ 
\\ 
PJ
\end{array}
$ & $
\begin{array}{c}
EK \\ 
0.55 \\ 
\\ 
4.3 \\ 
\\ 
0.8
\end{array}
\begin{array}{cc}
BH & ALL \\ 
2.15,1.05 & 2.15,1.1 \\ 
&  \\ 
11,6.0 & 11,6.6 \\ 
&  \\ 
3.0,1.45 & 3.0,1.55
\end{array}
$ & $
\begin{array}{cc}
BH & ALL \\ 
1.55,0.75 & 1.55,0.85 \\ 
&  \\ 
7.5,4.0 & 7.9,5.15 \\ 
&  \\ 
2.1,1.05 & 2.15,1.15
\end{array}
$ \\ \hline
\end{tabular}
\end{table}
\end{center}
\end{widetext}

Our results are summarized in Fig. 3, where we show the minimum value of the
scale of gravity allowed by RICE data at $95\%$ C.L. For a given case, the
region \textit{below} the curve is excluded. The points on the curves are
obtained by setting the value of $M_{BH0}$, and then varying $M_{D}$ with $%
d=6$ until the predicted number of events falls below 3, consistent with the
RICE result of zero events on zero background at $95\%$ C.L. The lower bound
on $M_{D}$ is plotted against $M_{BH0}/M_{D}$, the ratio of a minimum
invariant mass required for black hole formation to the scale of gravity. In
setting the limit curves, the SM, LSG unitarized graviton exchange, EK, and
the LSG black hole production cross sections, BH and $BH_{inel}$, are all
included in determining event rates and, consequently, the bound. The top
curve for each flux model represents the minimum $M_{D}$ when the simple
''black disk'' model for the black hole formation cross section is used; the
bottom curve indicates the lower limit when impact parameter dependence of
the apparent horizon is included as estimated in \cite{impactpar} and
implemented by \cite{afgs}. The impact of the uncertainty in correct
cosmogenic flux model and model of BH cross section on the bound of $M_{D}$
is seen at a glance in Fig. 3. Fixing $M_{BH0}/M_{D}$, the range of possible
bounds on $M_{D}$ follows by comparing the lowest curve to the highest.

To contrast the above results to the those from nuclei-generated cosmogenic
neutrino flux estimates, we calculated bounds for the case $M_{BH0}/M_{D}$ =
1 in ''small'' (cutoff at $10^{6.5}PeV$) and ''large''(cutoff at $10^{7.5}PeV
$) versions of the iron-based flux model from the first reference in \cite
{nunucl}. Using the most pessimistic black hole production assumptions in
the ''small'' flux case and the most optimistic for the ''large'', the lower
bound range is $0.5TeV<M_{D}<1.0TeV$. Models based on lighter nuclei such as
oxygen or helium produce neutrino fluxes, and therefore $M_{D}$ limits, much
closer to those found from the models based on protons as the UHE cosmic
rays.

\begin{figure}[t]
\includegraphics[width=3.8in,angle=0]{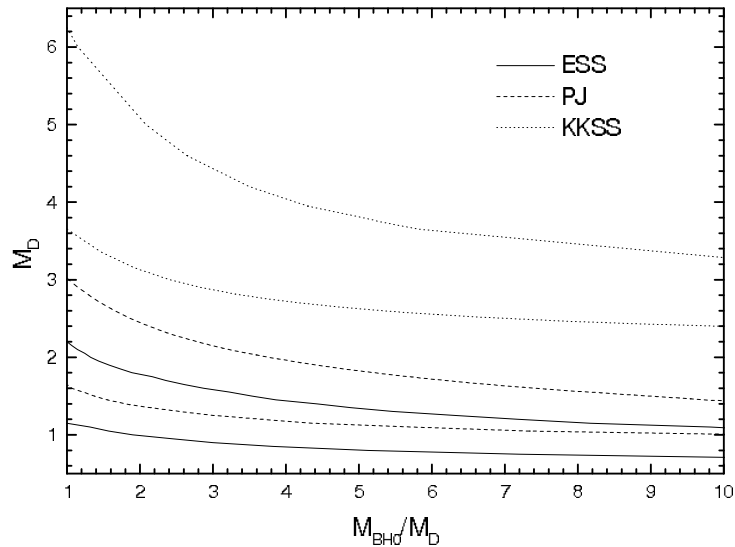} \vspace{0.0cm}
\par
\caption{Lower bounds on $M_{D}$ as a function of the ratio of the minimum
BH formation threshold, $M_{BH}$, to $M_{D}$. The upper curve for each flux
model is the lower bound when the naive, black disk model is used for the BH
cross section. The lower curve is the lower bound when the estimate of
impact parameter effects is included. ESS, KKSS (lower of the two KKSS
curves in Fig.3) and PJ refer to Refs. \protect\cite{ess}, \protect\cite
{kkss} and \protect\cite{pj}}
\label{fig:lsglim}
\end{figure}

\section{Discussion of results and Outlook}

The essence of our results is shown in Fig. 3, where the region below a
given curve shows the excluded region for which $M_{D}$ is too small, and
the LSG cross section too big, to be consistent with the RICE results \cite
{rice05}. As we comment further below, in connection with the information
contained in the tables, the conclusions depend heavily on the flux model
assumed and the treatment of the LSG interactions. It is clear that, should
events be observed, there are likely to be a number of different parameter
choices in the flux models and the cross section models that will reproduce
the observed events. This will still be true if a component of heavy nuclei
is present in the UHE cosmic rays, modifying the cosmogenic flux predictions.

Within the range of cosmogenic flux models we consider, Table III reveals
that at 95 \% C. L., RICE and the ESS flux model \cite{ess} rule out an LSG
model where the naive $\sigma _{BH}=\pi r_{S}^{2}$ cross section is assumed,
where $M_{BH}=3M_{D}$, and where $M_{D}<1.55$ $TeV$. This can be regarded as
a least lower bound on models with the naive ''black disk'' cross section
for the black hole formation within our analysis assumptions. The greatest
lower bound corresponds to that obtained with the largest flux model, and
the value is greater than 7.9 $TeV$ (11 $TeV$) when $M_{BH0}=3M_{D}$ ($M_{D}$%
). The KKSS maximal flux model \cite{kkss} is right at the borderline of our
95 \% C.L. constraint using the SM cross section\cite{rice05}, so the
corresponding LSG scale is imprecise. Basically there is no room for LSG if
this flux model is assumed.

If the estimates of the effects of non-zero impact parameter are included in
the manner proposed in \cite{impactpar}, and some of the collision energy is
lost to the BH formation process, then the BH formation cross sections
decrease and the event rates and corresponding bounds on the LSG scale
weaken. The nominal effects on the least lower bound, tied to the flux model 
\cite{ess}, and greatest lower bound, tied to KKSS \cite{kkss} are the
reductions to 0.85 $TeV$ and 5.15, respectively. The greatest lower bound of
all, 11$TeV$, occurs if the BH scale is equal to $M_{D}$ and the simple
black disk cross section is used with the maximum KKSS flux shown in Fig. 2.
Given the lack of precision in these considerations, we summarize the most
conservative bounds statement as an estimated range on the value of $M_{D}$
as lying in the range between roughly 0.9 $TeV$ and 10 $TeV$, as quoted in
the Abstract.

Next, suppose for the sake of argument that a given number of events were
actually observed in the RICE data. There are interesting degeneracies that
occur. For example, there is an approximate degeneracy between the EK $M_{D}$
= 1, 2 $TeV$ entries and $BH$ ($M_{D}$ = 1, 2 $TeV$ with $M_{BH0}$ = 3 $TeV$%
) entries for every flux model we used. The enhancements in event rates are
more than an order of magnitude compared to SM in every case, but the
interpretation, just within low scale gravity itself and with the flux
known, will be a challenge. The $EK$ calculation does not invoke black hole
formation, but for one scale choice $EK$ can yield the same signal
enhancement as $BH$ with a different scale choice.

When the extra uncertainties in flux are admitted, there is a sharp increase
in the possible degeneracies in parameter choices, leading to similar event
rates. For example, it is no surprise to find in Table III that the ESS flux
with BH1 parameter choices give the same RICE event rate predictions as the
PJ flux with the BH2 parameter set for the LSG model, with and without
inelasticity included. Similarly, the PJ flux with EK interactions alone
produce, as noted in Table III, the same bound on $M_{D}$ as the ESS flux
with ALL interactions included and inelasticity estimated for the BH
production case.

A number of ideas have been proposed for distinguishing SM from new physics
cross sections, and new physics cross sections from one another,
independently of flux; see for example Ref. \cite{prop} and references cited
there. Typically these ideas require angular information, at least number of
up-going events vs. down-going events, to "divide out" the flux effects. To
get discriminating power takes a sizable sample of events, which is a
challenge for all UHE telescope projects.

These remarks remind us that the field of UHE neutrino telescope physics and
astrophysics is still very young. All-in-all we are at a very complex, fluid
and fascinating threshold of discovery. Much more experimental and
theoretical work remains to be done.

\textbf{Acknowledgements} We thank Dave Besson and the RICE Collaboration
for discussions and encouragement. Comments on early versions of the
manuscript from Danny Marfatia are appreciated. John Ralston and Prasanta
Das participated in the early stages of this work, and we thank them for
their contributions. We appreciate the cooperation of D. Seckel, R.
Protheroe, D. Semikoz and A. Taylor, who provided us with data files from
their cosmogenic flux model calculations. This work was supported in part by
the Department of Energy High Energy Physics Division and by the NSF Office
of Polar Programs.

\end{document}